\documentstyle[prl,aps,psfig]{revtex}

\begin{document}


\title{A New Type of Electron  Nuclear-Spin Interaction from Resistively Detected NMR in the Fractional Quantum Hall Effect Regime}
\author{S. Kronm\"uller,  W. Dietsche and K. v. Klitzing}
\address{Max-Planck-Institut f\"ur Festk\"orperforschung, Stuttgart, Germany}
\author{G. Denninger} 
\address{2. Physikalisches Institut,Universit\"at Stuttgart, Stuttgart, Germany}
\author{W. Wegscheider and M. Bichler}
\address{Walter Schottky Institut, Technische Universit\"at M\"unchen, M\"unchen, Germany }
\date{\today}
\maketitle

\begin{abstract}

Two dimensional electron gases in narrow GaAs quantum wells show huge 
longitudinal resistance (HLR) values at certain fractional filling factors. 
Applying an RF field with frequencies corresponding  to the nuclear 
spin splittings of  $\rm ^{69}Ga$,  $\rm ^{71}Ga$ and $\rm ^{75}As$
leads to a substantial decreases of the HLR establishing a novel 
type of resistively detected NMR.
These resonances  are split into four sub lines each. Neither the number of sub lines nor the size 
of the splitting can be explained by established interaction mechanisms.

\pacs{73.40.Hm,73.20.Dx}       
\end{abstract}



Two-dimensional electron gases (2DEGs) with very high mobilities of the
electrons can be formed in quantum wells and heterostructures based on the
$\rm GaAs/Al_xGa_{1-x}As$ system. If such a 2DEG is subjected to an intense
perpendicular magnetic field at very low temperatures, it shows the
integer \cite{klitzing} and the fractional \cite{tsui} quantum Hall effects 
at integer and fractional filling factors of one or more Landau levels.
The signature of both types of quantum Hall effects is the quantization of the
Hall resistance and the vanishing of the longitudinal resistance. 
Recently, however, huge longitudinal resistance maxima (HLR) have been observed 
at fractional filling factors between $\rm \frac{1}{2}$ and 1 \cite{kron}.
The HLR is only found in samples which have a reduced well thickness (15 nm, \cite{kron1})
as compared to the conventional ones.
As an example,
figure 1 shows longitudinal resistance measurements on a sample similar to the one 
used in  \cite{kron} for two different carrier densities (dotted and dashed line) 
at a temperature of 0.35 K.
Here, the magnetic field is swept at a rate of  $\rm 0.7\,T/min$ 
and the applied source
drain current is  $\rm 100\,nA$. The width of the sample is 80 $\rm \mu m$ and the voltage probes
are 80 $\rm \mu m$ apart.
For both carrier densities 
a very regular behavior is seen. 
At integer filling factors the resistance vanishes completely  and at filling factor 
$\rm \nu = \frac{2}{3}$ one finds a clear minimum. 
However, if the
sweep rate of the magnetic field is drastically reduced to 
$\rm 0.002\,T/min$, a huge maximum in the longitudinal resistance  
(solid lines) 
is observed at $\rm \nu = \frac{2}{3}$  for both carrier densities.
The size of the HLR is maximal at a current density of approximately
  $\rm 0.6 \,mA/m$.
The HLR vanishes in tilted magnetic fields, indicating that the electron 
spin polarization plays an  important role for the HLR.
Similar maxima are also reported at other fractional 
filling factors \cite{kron}, but in this paper we want to concentrate on
the HLR at $\rm \nu = \frac{2}{3}$ at 0.35K.

The  HLR develops with a time constant of about 15 min.
These very long times are typical 
for relaxation effects of the nuclear spin system \cite{dobers,berg}.
The only direct way to demonstrate an involvement of the nuclear spins in the
HLR is a nuclear magnetic resonance (NMR)
\cite{flinn,krapf,barrett,guerrier} experiment, because it allows direct
modification of the nuclear polarization. 
In this Letter we report on experiments where radio frequency is irradiated on a sample in
the HLR state and a drastic reduction of the resistance values is observed whenever the nuclei are
in resonance. This is to our knowledge the clearest form of a resistively detected NMR in a solid state 
system. Moreover we report that the NMR resonances are split into four sub lines which can
neither be explained by dipole-dipole interaction between neighboring nuclei nor by hyperfine
interaction with the electrons. This indicates that the HLR is indeed a novel fractional state.

There have been only a few experiments where the interaction between
electrons and nuclear spins in GaAs quantum wells has
been probed. Effects of the nuclear polarization on the electron
transport have been
observed after a non equilibrium electron spin distribution was first
produced by ESR radiation  \cite{dobers,berg} or by tunneling
between different spin polarized Landau levels  \cite{kane,wald,dixon} which was then
transfered into the nuclear system via the hyperfine interaction.
Alternatively it is possible to pump the nuclear polarization optically
and observe the NMR either inductively \cite{barrett} or optically \cite{flinn,krapf,guerrier,kukushkin}.
The HLR seems to represent a completely different situation
since the nuclear spin polarization occurs without special experimental 
preparation.


In our NMR experiment we measure the longitudinal resistance
of a  modulation doped $\rm 15\,nm$ thick GaAs quantum well embedded in 
$ \rm Al_{0.3}Ga_{0.7}As$ barriers. The
carrier density is about  $\rm 1.3 \!\cdot\! 10^{11} \,cm^{-2} $ and
the mobility  $\rm 1.8 \!\cdot\! 10^{6}\, cm^2/Vs$ after
 illumination  with a LED. The measurements are  performed  in a $^3$He
 bath cryostat at $\rm 0.35\, K$  using an ac lock-in technique with a 
modulation frequency of 23 Hz. To create a  radio frequency (RF) 
magnetic-field perpendicular 
to the static magnetic field we put a wire loop around our sample to which
RF is applied (inset fig. 1). The loop is mounted so 
that its normal direction  is perpendicular
to the static magnetic field. 
We performed the NMR experiments on two different Hall bar samples which were
$\rm 800\, \mu m$ and  $\rm 80\, \mu m$   wide, using
a source drain current of $\rm 400\,nA$ and  $\rm 50\,nA$ respectively. 
During the experiment the HLR maximum is allowed to 
develop at constant magnetic field until it reaches its 
peak value. Then the RF is applied and 
 its frequency is swept over the range at which the nuclear resonances
are expected while the resistance is monitored as a function of the RF
 frequency.

Results are shown in figure 2
where the longitudinal resistance is plotted as a function of the RF 
frequency for three different carrier densities.
Minima are indeed found at frequencies corresponding
to the nuclear resonance frequencies of $\rm ^{69}Ga$ \cite{bruker}.
The resonance frequencies shift 
because the HLR occurs at different magnetic fields for different densities.
These traces are the first observation of NMR directly in the longitudinal 
resistance  of a 2DEG.

Similar NMR resonances are also observed at the expected respective frequencies 
for the other $\rm ^{71}Ga$ isotope and for
$\rm ^{75}As$. At all resonances the HLR is approximately 
reduced by 5 to 10\%.
The HLR recovers fully after leaving the resonance region to the high or 
low frequency side.
We take this as  clear evidence that the HLR is connected with 
a polarization of the nuclei, because continuous irradiation of the nuclear system
with a NMR resonance frequency saturates this transition leading to equal population of 
the respective spin levels. During this process the nuclear polarization is reduced.

No resonance signal is observed for Al therefore one can conclude 
that the relevant nuclear spin polarization is indeed created in the quantum well only.
The observed NMR minima cannot be due to resonant heating via the nuclei since
the energy absorption is vanishingly small due to the long $\rm T_1$ relaxation time. 
The line shape of the resonance depends
on the sweep rate of the RF-frequency (inset figure \ref{Figure2}). 
The slow sweep shows a very symmetric resonance so that one can assume the
system is in equilibrium at all times, which 
is not the case for the fast sweeps. 
For the fast sweep rates a sharp drop of the HLR is observed 
when approaching the
resonance. 
When leaving the resonance the HLR recovers on a timescale of several minutes,
which is similar to the time scale needed for the HLR to develop in the first
place.

If one uses a smaller sized sample then the effect of the inhomogeneity of the
magnetic field, leading to inhomogeneous broadening of the NMR line, is reduced. 
Figure  \ref{Figure0} shows the resonance lines of all three
isotopes $\rm ^{75}As$, $\rm ^{69}Ga$ and $\rm ^{71}Ga$ for the $\rm 80 \; \mu m $ wide  sample.
With the smaller  sample we indeed observe 
that the NMR resonances split into four sub lines.
The splitting is most pronounced for the As resonance lines.
The separation between the respective four lines is nearly equidistant
for all three isotopes and we find as  average values
of the splitting 30 kHz for $\rm ^{75}As$, 14.5 kHz for $\rm^{69}Ga$
and 10 kHz for  $\rm^{71}Ga$. 

The splitting into four sub lines is very surprising. The three nuclei
have a spin of I=3/2 which corresponds to a fourfold degenerate nuclear spin ground state,
which splits in a magnetic field by 
$\rm E_Z \;=\; \gamma_n \hbar B_0 m_I$ \cite{slichter}; where $\rm \gamma_n$
is the gyromagnetic ratio, $\rm B_0$ is the externally applied static magnetic field and $\rm m_I$
is the z-component of the nuclear spin.
Three but not four  different resonance frequencies would result if the  electric
quadrupole moment couples to an electric field gradient $\rm V_{zz}$.
Thus, the quadrupole moment cannot be responsible for the four resonance lines.

Another possibility to account for the splitting of nuclear resonance lines
is direct dipolar coupling between two neighboring nuclear spins. 
The coupling to an isolated second spin I=3/2 would indeed lead to a fourfold
splitting. However in a solid this
leads only to a broadening because there are several different species of neighboring spins
which have different distances from each other. Furthermore, the strength of the dipolar coupling 
is less than one kHz for the nuclear distances in GaAs, which is much too small.
A similar argument applies to the effect of unknown impurities, which would be 
statistically distributed and therefore can only lead to a broadening of the NMR lines.

The  nuclear moments can of course also interact with  the electronic system.
However, the standard treatment of the effect of quasi metallic electrons on
NMR leads only to a shift, the so called Knight Shift, but not to a splitting of the resonance lines.

Alternatively one could argue that the electrons possess an effective spin S=3/2,
which would lead to a fourfold split nuclear resonance via the hyperfine interaction.
The hyperfine interaction for the coupling between an s-electron with a nuclear spin
is given by $\rm H_{HF}\;=\;\frac{2}{3} \mu_0 g_0 \mu_B \hbar \gamma_n 
 \left| \Psi(0)\right|^2  \boldmath \vec{I}\cdot \vec{S} \unboldmath$ \cite{slichter,paget}, where
$\rm \mu_B$ is the Bohr magneton
, $\rm g_0$ is the g-factor of the free electron
, $\rm \boldmath \vec{I} \unboldmath$ is the nuclear spin
, $\rm \boldmath \vec{S} \unboldmath$ is the electron spin and
  $\rm \left|\Psi(0)\right|^2$ is the electron density at the nuclear site.
Introducing the values from literature \cite{bruker} leads to a hyperfine energy 
of 14660 MHz  per nucleus for $\rm ^{75}As$,
 12210 MHz for $\rm ^{69}Ga$ and 15514 MHz for $\rm^{71}Ga$.
Scaling these values with the electron densities in the GaAs quantum well
 with respect to the density in a metal, leads to hyperfine splittings 
of 27, 21, and 26 kHz for $\rm ^{75}As$, $\rm ^{69}Ga$ and $\rm^{71}Ga$ respectively. 
These splittings are of the experimentally observed
order of magnitude. However, the theoretical ratio of the hyperfine
splitting between the $\rm ^{69}Ga$ and the $\rm^{71}Ga$ isotope is 0.79 while the
experimental one is 1.4. This divergence between the two Ga isotopes is
unacceptable and rules out the hyperfine interaction with an effective
electron spin of 3/2 as the sole reason for the fourfold splitting of the
resonances.
Nevertheless, the order of magnitude of the observed splittings points
towards the hyperfine interaction. It is noteworthy that the ratios between
the experimentally observed values of the splittings can be quite well
described by scaling the theoretical hyperfine interaction with the natural abundance
of the isotopes, which is 0.61 for $\rm ^{69}Ga$ and 0.39 for $\rm ^{71}Ga$.
This would increase the theoretical ratio to 1.2, which is close to the experimental value
of 1.4. 
This could suggest that the splitting is caused
by the dipole interaction to other nuclei after all, but that the coupling
strength is enhanced by the hyperfine interaction to the electrons. Since
the other nuclei have I=3/2, this coupling has the potential to lead to a
fourfold splitting if the broadening from having many neighboring isotopes
at different distances is not effective. Such a mechanism would, however,
be completely novel and 
would amount to a new  correlated phase between the electrons and the nuclei.
At this time, however, the existence of such a phase is purely speculative.

The results of our investigation can be summarized as follows:
First, the resistance value of the HLR maxima drops if RF with frequencies
 corresponding to the splitting of the nuclear spins, 
is present. Since the RF irradiation leads to the
saturation of the nuclear transitions 
and to a reduction of the nuclear polarization, we conclude 
that the HLR is caused or stabilized by a
nuclear magnetic polarization. This polarization can only build up dynamically by the current
flow, which indicates that the electronic transport is connected with spin flip 
processes.

Second, whatever the exact nature of the HLR maximum is, it must take place
in the GaAs quantum well only. Because otherwise we would also have
observed nuclear magnetic resonances corresponding to the Al nuclei. The Al
nuclei are absent in the well but are present in the surrounding AlGaAs
barrier material.

Third, the splitting of the resonance lines into four lines is unexpected and
not yet understood. At this time one can only say that the size of the splitting  scales
 approximately with 
the product of the hyperfine interaction and the natural abundance of the
respective isotope and, that the size of the splitting is in the range of the
expected splitting for the hyperfine interaction. 
To our knowledge, none of the commonly discussed interaction mechanisms can lead to 
the observed fourfold splitting.

In conclusion we have found that the electron transport in the HLR state must be 
related to a dynamic polarization of the nuclei. This fact gives the rare opportunity
to detect NMR resistively in this state. The fine structure of the NMR  lines points to 
an unusual correlation between electrons and nuclei.

We acknowledge many discussions with T. Chakraborty, R. Gerhardts , and M.
Mehring. We received important experimental help from J. Weis and U. Wilhelm.



\begin{figure}
\caption{ Longitudinal resistance for two different carrier densities. The longitudinal resistance maximum 
 (HLR) at filling factor 2/3 is clearly developed for the
two carrier densities at the slow sweep rates ($\rm 0.002\; T/min$) of the magnetic fields. The inset shows the experimental setup. }
\label{Figure1}
\end{figure}

\begin{figure}
\caption{ The resistively detected NMR signal of $\rm ^{69}Ga$. A current of 400 nA is passed through an 800 $\rm \mu m$ wide  2DEG sample and the HLR is at its peak value. A voltage drop of approximately 10\% in the longitudinal voltage is observed when the RF is in resonance with the nuclei. The figure shows the resonance line for three different magnetic fields, which means three different carrier densities. The insert is the   $\rm ^{69}Ga$ resonance for different sweep rates, 1kHz/3s and 1kHz/60s,  of the RF frequency. The resonance is symmetric only if the frequency is swept slow enough.}
\label{Figure2}
\end{figure}

\begin{figure}
\caption{The NMR resonance line for $\rm ^{75}As$, $\rm ^{69}Ga$ and $\rm ^{71}Ga$ measured on the $\rm 80 \; \mu m $ wide sample. The tick marks are in 10 kHz steps.
A fourfold substructure is visible (arrows) for all three isotopes.}
\label{Figure0}
\end{figure}


\begin{references}
\bibitem{klitzing}K. von Klitzing, G. Dorda and M. Pepper, Phys. Rev. Lett. {\bf 45}, 494 (1980).
\bibitem{tsui}D.C. Tsui, H.L. Stormer and A.C. Gossard,  Phys. Rev. Lett. {\bf 48}, 1559 (1982).
\bibitem{kron}S. Kronm\"uller, W. Dietsche, J. Weis,  K. von Klitzing, W. Wegscheider and M. Bichler, Phys. Rev. Lett. {\bf 81 }, 2526 (1998).
\bibitem{kron1}We observe the same HLR phenomena also in 14 nm quantum wells.
\bibitem{dobers}M. Dobers, K. v. Klitzing, J. Schneider, G. Weimann and K. Ploog, Phys. Rev.Lett. \ {\bf 61}, 1650 (1988).
\bibitem{berg}A. Berg, M. Dobers, R.R. Gerhardts and K. v. Klitzing, Phys. Rev.Lett. \ {\bf 64}, 2563 (1990).
\bibitem{flinn}G.P. Flinn, R.T. Harley, M.J. Snelling, A.C. Tropper and T.M. Kerr,  Semicond. Sci. Technol. {\bf 5}, 533 (1990).
\bibitem{krapf}M. Krapf, G. Denninger, H. Pascher, G. Weimann and W. Schlapp, Solid. State. Comm. {\bf 78}, 459 (1991).
\bibitem{barrett}S.E. Barrett, R. Tycko, L.N. Pfeiffer and  K.W. West, Phys. Rev. Lett. {\bf 72}, 1368 (1994).
\bibitem{guerrier}D.J. Guerrier and  R.T. Harley, Appl. Phys. Lett. {\bf 70}, 1739 (1997).
\bibitem{kane}B.E. Kane, L.N. Pfeiffer and K.W. West, Phys. Rev. B {\bf 46}, 7264 (1992).
\bibitem{wald}K.R. Wald, L.P. Kouwenhoven, P.L. McEuen, N.C. van der Vaart and C.T. Foxon, Phys. Rev. Lett. {\bf 73}, 1011 (1994).
\bibitem{dixon}D.C. Dixon, K.R. Wald, P.L. McEuen and M.R. Melloch, Phys. Rev. B {\bf 56}, 4743 (1997).
\bibitem{kukushkin}I.V. Kukushkin, K. von Klitzing and K. Eberl, Physica E {\bf 1}, 21, (1997) and I.V. Kukushkin to be published
\bibitem{bruker}Standard NMR tables for instance Bruker Almanac, (Bruker GmbH, Karlsruhe, 1998).
\bibitem{slichter}C.P. Slichter, Principles of Magnetic Resonance, Springer Verlag, Berlin (1990).
\bibitem{paget}D. Paget, G. Lampel, B. Sapoval and V.I. Safarov, Phys. Rev. B {\bf 15}, 5780 (1977).
\end{references}
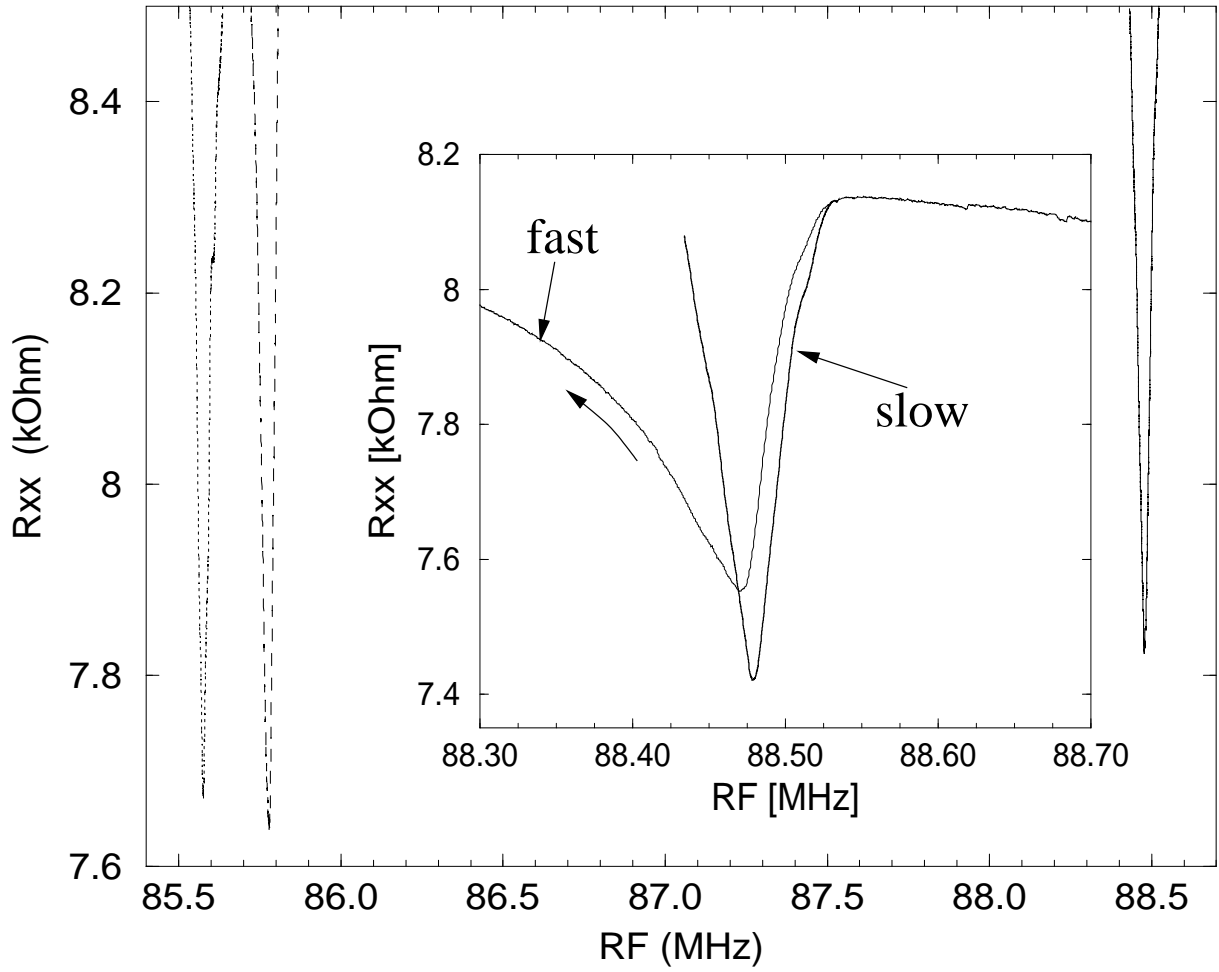
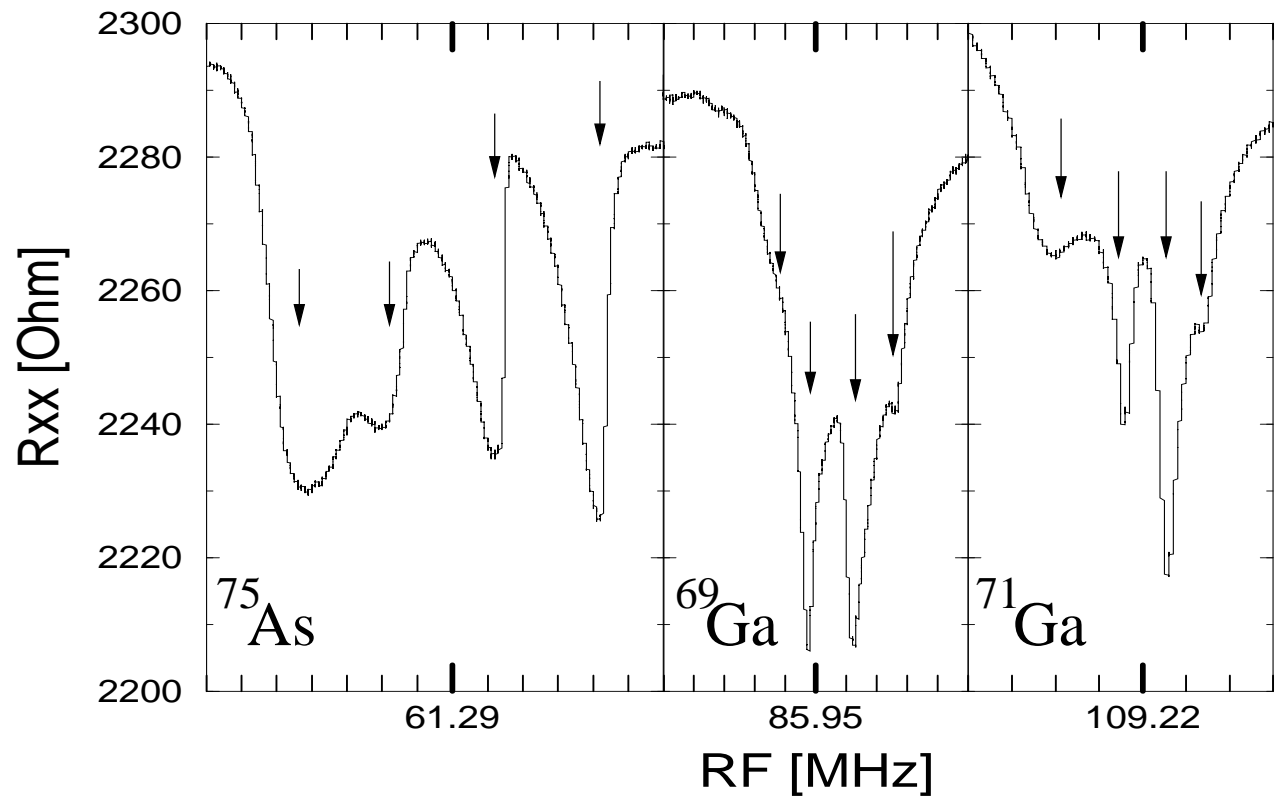
\end{document}